\documentclass[twocolumn,showpacs,preprintnumbers,amsmath,amssymb]{revtex4}

\usepackage{graphicx}
\usepackage{dcolumn}
\usepackage{bm}

\begin{document}


\title{Flow fields in soap films: relating surface viscosity and film thickness}

\author{V.~Prasad}
\author{Eric R.~Weeks}%
\affiliation{ Department of Physics, Emory University, Atlanta, GA
30322 }

\date{\today}

\begin{abstract}
We follow the diffusive motion of colloidal particles in soap films
with varying $h/d$, where $h$ is the thickness of the film and $d$
the diameter of the particles. The hydrodynamics of these films are
determined by looking at the correlated motion of pairs of particles
as a function of separation $R$. The Trapeznikov approximation [A.
A. Trapeznikov, \emph{PICSA} (1957)] is used to model soap films as
an effective interface in contact with bulk air phases, that behaves
as a 2D fluid. The flow fields determined from correlated particle
motions show excellent agreement with what is expected for the
theory of 2D fluids for all our films where $0.6 \leq h/d \leq
14.3$, with the surface viscosity matching that predicted by
Trapeznikov.  However, for thicker films with $h/d > 7 \pm 3$,
single particle motion is faster than expected.  Additionally,
while the flow fields still match those expected for 2D fluids,
the parameters of these flow fields change markedly for thick
films.  Our results
indicate a transition from 2D to 3D fluid-like behavior occurs at
this value of $h/d$.

\end{abstract}

\maketitle

\section{Introduction}
The motion of a particle in a viscous fluid causes a flow field to
be created, that is, fluid mass is displaced in a very specific
manner around the particle. The nature of this field depends
sensitively on the geometry, boundary conditions and dimensionality
of the system in question. For instance, flow fields in 3D decay as
$1/R$ in an unbounded fluid, where $R$ is the distance from the
localized perturbation. On the other hand, in a 2D fluid such as a
thin film, flow fields decay logarithmically with distance
\cite{stone1,levine1,fischer1}. One example of a 2D system
exhibiting such long-range behavior under certain specific
circumstances is that of soap films
\cite{diLeonardo1,Cheung1,Mathe1}. A soap film in its simplest form
consists of a thin fluid layer of thickness $h$ buffered from air
phases above and below it by surfactant layers. The fluid layer has
a 3D viscosity $\eta_{\text{bulk}}$, and the surfactant layers a 2D
\emph{surface} viscosity $\eta_{\text{int}}$ \cite{wu2,ecke1}.
However, because of the finite thickness of the fluid layer, and the
2D nature of the surfactant layers, it is unclear whether a soap
film should be regarded as a 2D or 3D fluid for arbitrary film
thickness, bulk viscosity and surface viscosity.

Two particle microrheology \cite{crocker1,levine2} provides a
powerful technique to determine the flow fields in soap films. This
technique has been used with great success to characterize the
rheological and flow properties of 3D systems such as biomaterials
\cite{gardel1}, polymer solutions \cite{starrs1,dasgupta1} and
biological cells \cite{lau1}. To a lesser extent, it has also been
used to determine the surface rheological properties of 2D fluids
such as protein monolayers at an air water interface \cite{prasad1}.
Briefly, this technique looks at the correlated motions of particles
in the system of interest. When a particle undergoes motion in the
system, either by thermal excitation \cite{crocker1,prasad1} or by
an external perturbation \cite{starrs1,diLeonardo1}, it creates a
flow field in the surrounding medium. This flow field affects the
motion of other particles in its vicinity. Therefore, by measuring
the correlated motions of pairs of particles as a function of their
separation $R$, the flow field in the system can be determined.

In this manuscript, we look at the flow fields in soap films of
different thickness $h$ and bulk viscosity $\eta_{\text{bulk}}$,
while keeping the surface viscosity $\eta_{\text{int}}$ constant. We
embed probe particles of size $d$ in these soap films, and are able
to vary the dimensionless parameter $h/d$ by over an order of
magnitude, from $h/d=0.6-14.3$ \cite{prasad3}. In order to account
for the effects of the surfactant-laden interfaces and the thin
fluid layer, we use the Trapeznikov approximation
\cite{Trapeznikov1} that models the soap film as an effective
interface with an effective surface viscosity $\eta_{T} =
\eta_{\text{bulk}}h+2\eta_{\text{int}}$. The flow fields in these
films are then mapped as a function of $R$, and compared to
theoretical models of soap film flow. Excellent agreement is
obtained between the theoretical models and the flow fields in
\emph{all} soap films, irrespective of the parameter $h/d$, using
$\eta_T$. However, we find that single-particle motion is faster
than expected for thicker films with $h/d > 7 \pm 3$. This leads us
to state that the hydrodynamics of the soap films transition from 2D
to 3D-like behavior at this particular value of $h/d$.

\section{Experiments}

\subsection{Preparation of soap films}\label{expt-1}
Soap films are prepared from mixtures of water, glycerol and
surfactants that stabilize the interfaces. By changing the ratio of
water and glycerol, the viscosity $\eta_\text{{bulk}}$ of the soap
solution can be controlled. The surfactants used in this study are
obtained from a commercially available dishwashing detergent brand
Dawn. Known quantities of this detergent are added to the
water/glycerol mixture to create the soap solutions. The chemical
formulation of Dawn is proprietary and hence cannot be determined;
however, for all the soap films in this study we use the same amount
of dishwashing detergent (2$\%$ by weight) to ensure consistency.

Fluorescent polystyrene spheres (Molecular Probes, carboxylate
modified, $d=210$, 500 nm) are added to the soap solutions. The soap
films are created by dipping a circular stainless steel frame of
diameter 1 mm into the solutions and drawing it out gently. The
frame is then enclosed in a chamber designed to maintain relative
humidity and minimize convective drift. The particles are then
imaged with fluorescence microscopy to determine their motion in
real space and real time.

\subsection{Particle tracking by fluorescence microscopy}
\label{tracking} The tracer particles are imaged in a fluorescent
microscope at a frame rate of 30 Hz, with a 20$\times$ objective
(numerical aperture = 0.4, resolution = 465 nm/pixel) used for large
particles ($d=500$ nm), and a 40$\times$ objective (numerical
aperture = 0.55, resolution = 233 nm/pixel) for smaller ($d=210$ nm)
particles. For each sample, short movies of duration $\sim 30$ s are
recorded with a CCD camera that has a 640 $\times$ 486 pixel
resolution, with hundreds of particles lying within the field of
view.
The movies are later analyzed by particle tracking to obtain the
positions of the tracers \cite{crocker96}.
From the particle positions, we determine
their vector displacements by the relation $\Delta
r(t,\tau)=r(t+\tau)-r(t)$, where $t$ is the absolute time and $\tau$
is the lag time. Any global motion is subtracted from these vector
displacements to minimize the effects of convective drift caused by
the air phases that contact the soap film. These vector
displacements are then used to determine the mean square
displacement (MSD), $\langle\Delta r^{2}(\tau)\rangle$, where the
average is performed over all particles and all times $t$.
Correlated motions of particles \cite{crocker1,mason2} are also
determined by looking at the products of particle displacements,
which we describe in Sec.~\ref{theory} in greater detail.

\subsection{Determination of soap film thickness}\label{expt-2}

The thickness of the soap films in this study range from $h \sim
300$ nm - 3 $\mu$m, close to the wavelength of visible light. This
make the spectroscopic technique of optical interference the
most viable option for determining the thickness of these films.
Immediately after taking each movie, the film is transferred to
a spectrophotometer and its thickness $h$ determined from the
transmitted intensity \cite{huibers1}.  We briefly describe the
details of this technique: Two light rays of the same wavelength
passing through a thin film will interfere with each other. This
interference will be constructive or destructive, depending on
whether one light ray has traveled an integer or half-integer
multiple of the wavelength with respect to the other ray. The
light transmitted through the film will have a minimum (or the
absorption will have a maximum) when
\begin{equation}
(2n\, \text{cos}\, \theta)\cdot h=(m+1/2)\lambda
\end{equation}
where $n$ is its index of refraction of the film, $\lambda$ is the
wavelength of light, $\theta$ is the angle of incidence (typically,
$\theta=90^{\text{o}})$, and $m$ is a non-negative integer. If
$\lambda_{pk,1}$ and $\lambda_{pk,2}$ are two successive maxima (or
minima) in the transmitted light then the thickness of the film can
be easily determined by the relation \cite{huibers1}
\begin{eqnarray}
h=\frac{1}{2n}[\frac{1}{\lambda_{pk,1}}-\frac{1}{\lambda_{pk,2}}]^{-1}.
\label{thickness-interference}
\end{eqnarray}

\begin{figure} [tbhp]
\includegraphics[scale=0.36]{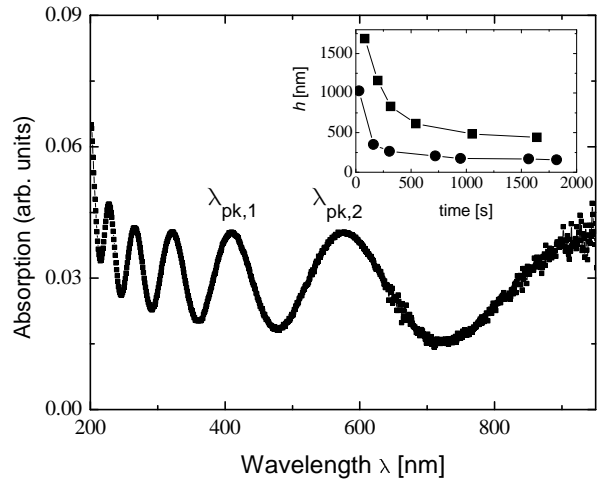}
\caption{\label{fig:thick} Absorption spectrum of a soap film (50:50
water/glycerol mixture, 2$\%$ Dawn by weight) as a function of
normally incident wavelength of light. From the peaks in this
spectrum and Eqn.~\ref{thickness-interference}, the thickness of the
film is inferred to be $h= 510$ nm. Inset: The time dependence of
$h$ for two different soap films
prepared with 60:40 water/glycerol ratio and $2 \%$ concentration of Dawn.}
\end{figure}

Figure~\ref{fig:thick} demonstrates how the thickness of a soap
film is determined in practice. It shows the absorption spectrum for
a soap film (50/50 water-glycerol mixture with 2$\%$ Dawn) that has
been placed in a UV-Vis spectrophotometer (Agilent Technologies,
Santa Clara, CA). The spectrum shows multiple maxima and minima because
of constructive and destructive interference as light rays traverse
through the soap film. By substituting the values of the two
successive maxima $\lambda_{pk,1} = 409$ nm and $\lambda_{pk,2}=576$
nm in Eqn.~\ref{thickness-interference} and $n=1.4$ for a 50:50
water glycerol mixture, the thickness $h =504$ nm of the soap film
can be estimated. We then average the value of $h$ obtained by
repeating this process for all the successive peaks observed in the
spectrum, giving us $h = 510 \pm 10$ nm for this particular soap
film.

The inset to Fig.~\ref{fig:thick} shows a time series of film
thickness for two soap films comprised of a 60:40 water-glycerol
mixture with $2 \%$ Dawn as surfactant. It is evident from the
figure that both films thin rapidly over an initial time scale of
$\sim$ 500 s, but rapidly equilibrate to a quasi-steady state over
longer timescales where relatively small changes are seen in the
thickness. Care is taken in our measurements to ensure that
the particle trajectories are recorded after this initial transient
period. This has the added advantage that convective drift of the
tracers is also substantially reduced beyond 500 s, resulting in
more accurate measurements of particle motions.

\section{Theory} \label{theory}
\begin{figure} [tbhp]
\includegraphics[scale=0.45]{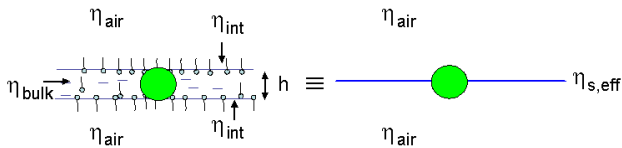}
\caption{\label{fig:cartoon2} Schematic of the Trapeznikov
approximation, where the entire soap film is approximated as a
single interface in contact with bulk air phases. Reproduced with
permission from \cite{prasad3}.}
\end{figure}

A soap film is considered ``thin'' when the thickness $h$ of the
film is comparable to the particle size $d$. In fact, it has been
shown that thin films behave as a 2D fluid \cite{prasad3,Cheung1}.
This assumption can be justified by modeling the soap film, with its
two interfaces and a thin fluid layer, as a single effective
interface in contact with two bulk air phases \cite{Trapeznikov1}
(see Fig.~\ref{fig:cartoon2}). The effective interface then has an
effective surface viscosity, $\eta_T$, which is given
by

\begin{equation}
\label{trapeqn}
\eta_T=\eta_{\text{bulk}}h+2\eta_{\text{int}}
\end{equation}
where $\eta_{\text{int}}$ is the surface viscosity of the two
surfactant-laden interfaces. One immediate consequence of this
approximation is that a tracer particle in a soap film must diffuse
as though embedded in an interface in contact with bulk air phases.
According to Saffman \cite{saffman1}, this diffusion follows the
equation
\begin{eqnarray}
\label{Saffman} \langle\Delta r^{2}\rangle =
\frac{k_{B}T}{\pi\eta_{S}}[\text{ln}(\frac{2\eta_{S}}{\eta_{\text{air}}d})-\gamma_{E}]\tau
\end{eqnarray}
where $\eta_{\text{air}}$ is the viscosity of the air phase and
$\gamma_E=0.577$ is Euler's constant.  It is expected that $\eta_S
= \eta_T$ and indeed this has been found to be true for thin films
\cite{prasad3}; the limits of this for thick films will be discussed
in Sec.~\ref{results}.

The other consequence of the Trapeznikov approximation is that the
hydrodynamics of a soap film must mimic that of a 2D fluid. To probe
the hydrodynamics, we look at the correlated motions of particles
embedded in the soap film. Similar to the treatment in
\cite{diLeonardo1}, we measure correlated and anti-correlated
coupled displacements of particles to determine the `eigenmodes' of
particle motion in a 2D fluid.  We give here a brief description
of the method used by di Leonardo \emph{et al.} \cite{diLeonardo1}.
They actively perturbed pairs of particles by means of optical
tweezers in a soap film.  The displacements of the particles from
their mean particle positions are then determined. These
displacements are related to strength of the trap and the
mobility of the particles. The mobility of each particle is affected
by the presence of the other particle, and therefore contains
information about the nature of the flow fields in the soap film.
The two dimensional Stokes equation can then be solved and the
resulting motions of the particles decomposed into eigenmodes of
motion. There are four such eigenmodes in 2D, given by
$\lambda_{x\pm}$ and $\lambda_{y\pm}$ where $x$, $y$ represent
motion parallel and perpendicular to the lines joining the centers
of the particles and the $\pm$ represent rigid motions and relative
displacements respectively. These mobilities are given by the
equation
\begin{eqnarray}
\label{diLeonardo-theory} \lambda_{x\pm} =
b[1\pm\frac{1}{4\pi\eta_{\text{bulk}}hb}
\text{ln}(\frac{L}{R})]\nonumber\\
\lambda_{y\pm} = b[1\pm\frac{1}{4\pi\eta_{\text{bulk}}hb}
(\text{ln}(\frac{L}{R})-1)]
\end{eqnarray}
where $R$ is the separation between the particles, $L$ is a
characteristic length scale and $b$ has units of mobility
(kg$^{-1}$s). This length scale $L$ demands some explanation; flow
fields in 2D are long-ranged and tend to diverge because of the
presence of the logarithmic term in Eqn.~\ref{diLeonardo-theory}.
This divergence is cut-off due to the presence of a length scale
$L$, which has many possible origins. Some of these include finite
size of the film, inertial effects, and viscous drag on the
interfaces from the surrounding bulk fluid phases (air). In the
subsequent sections, we will look at a series of soap films with
different material parameters ($\eta_{\text{bulk}}$, $h$) and
discuss in detail the origin of this length scale $L$. A detailed
derivation of Eqn.~\ref{diLeonardo-theory} can also be found in
reference \cite{diLeonardo1}. It is clear from
Eqn.~\ref{diLeonardo-theory} that the mobilities of the particles
split around a mean mobility $b$, with rigid motions ($+$) being
favored and relative displacements ($-$) being opposed. The
hydrodynamic interactions between the particles are also governed by
the fluid layer of the film, shown by the appearance of
$\eta_{\text{bulk}}h$ in the equation.

\begin{figure} [tbhp]
\includegraphics[scale=0.55]{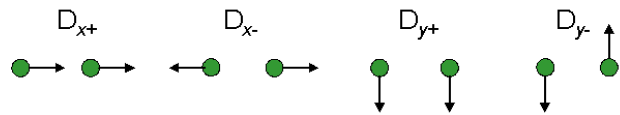}
\caption{\label{fig:cartoon} Coupled motions of particles embedded
in the soap films. There are 4 possible eigenmodes associated with
this coupling, correlated and anti-correlated motion parallel
($D_{x\pm}$) and perpendicular ($D_{y\pm}$) to the lines joining the
centers of the particles.}
\end{figure}

Our approach to determine the flow fields in soap films, while
analogous to di Leonardo's approach, has some important differences.
We measure the correlated \emph{thermal} motions of pairs of
particles embedded in the soap films. There are four such eigenmodes
in 2D, represented by $D_{x\pm}$ and $D_{y\pm}$ which correspond to
the longitudinal and transverse components of coupled motion, with
the $\pm$ representing correlated and anti-correlated motion
respectively (refer Fig.~\ref{fig:cartoon}). The correlation
functions are given by \cite{diamant1,diamant2}
\begin{eqnarray}D_{x\pm}(R,\tau)=\langle\frac{1}{2}[\Delta r_{x}^{i}(\tau)\pm\Delta
r_{x}^{j}(\tau)]^2 \delta (R-R^{ij})\rangle_{i\neq j}\nonumber\\
D_{y\pm}(R,\tau)=\langle\frac{1}{2}[\Delta r_{y}^{i}(\tau)\pm\Delta
r_{y}^{j}(\tau)]^2 \delta (R-R^{ij})\rangle_{i\neq j}
\end{eqnarray}
where $i$, $j$ are particle indices, the subscripts $x$ and $y$
represent motion parallel and perpendicular to the line joining the
centers of particles and $R^{ij}$ is the separation between
particles $i$ and $j$. The average is performed over all possible
pairs of particles with a given separation $R$. This has the
advantage of averaging out correlations from other particles that
are not part of the pair. We are therefore confident that many-body
effects due to the presence of other particles are minimized in our
correlation functions. Similar to \cite{prasad1}, we observe that
$D_{x\pm}$, $D_{y\pm} \sim \tau$ which enables the estimation of
four $\tau$-independent quantities $\langle
D_{x\pm}/\tau\rangle_{\tau}$ and $\langle
D_{y\pm}/\tau\rangle_{\tau}$ depending only on $R$ and having units
of a diffusion constant. In this manuscript, we shall discuss only
these $\tau$-independent quantities.

These $\tau$-independent correlation functions have a physical
interpretation; for instance, $\langle D_{x+}/2\tau\rangle_{\tau}$
is the diffusion constant of the center of mass of the particle
pairs along the line joining their centers, while $\langle
D_{x-}/2\tau\rangle_{\tau}$ is the diffusion constant of the
interparticle separation. Our correlation functions can then be
trivially related to di Leonardo's eigenmobilities by the relation
\begin{eqnarray}
\label{Leonardo-Haim}
\langle D_{x\pm}/\tau\rangle = 2k_{B}T \lambda_{x\pm}\nonumber\\
\langle D_{y\pm}/\tau \rangle = 2k_{B}T \lambda_{y\pm}
\end{eqnarray}

From Eqn.~\ref{diLeonardo-theory}, we can then derive theoretical
expressions for our thermally driven correlation functions which are
given by
\begin{eqnarray}
\label{Diamant}
\langle D_{x\pm}/\tau\rangle = B[1\pm C\text{ln}(\frac{L}{R})]\nonumber\\
\langle D_{y\pm}/\tau \rangle = B[1\pm C(\text{ln}(\frac{L}{R})-1)]
\end{eqnarray}
where $B=2k_{B}Tb$ has units of a diffusion constant and
$C=1/(4\pi\eta_{\text{bulk}}hb)$ is a non-dimensional constant. In
the subsequent section we explore the validity of these theoretical
expressions for a range of soap films with varying $h/d$ and
viscosity $\eta_{\text{bulk}}$ of the fluid layer comprising the
soap films.

\section{Flow fields in soap films}
\label{results}

\begin{figure} [tbhp]
\includegraphics[scale=0.38]{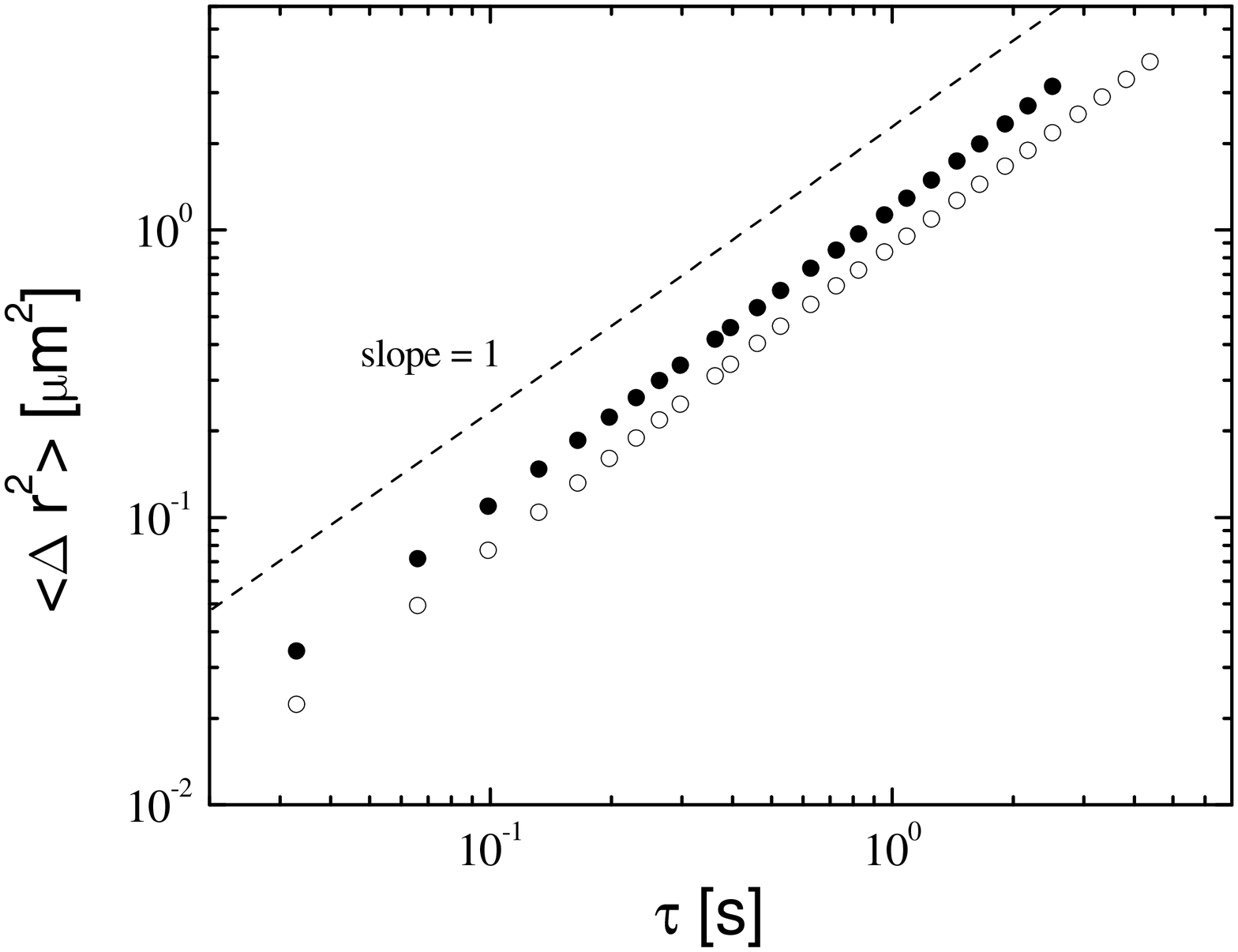}
\caption{\label{fig:msd-single} Mean square displacement (MSD) for a
soap film, sample f in Table \ref{table1} (solid circles) compared
to MSD in a bulk solution that comprises the fluid layer of the soap
film (open circles). Dashed line shows a slope of 1, indicating free
diffusion. The effective surface viscosity of the film can be
evaluated from Eqn.~\ref{Saffman}, and is given by
$\eta_{s,\text{eff}}=8.03$ nPa$\cdot$s.}
\end{figure}

Particle motions in the soap films are quantified by measurements of
the mean square displacement (MSD), $\langle\Delta r^{2}\rangle$,
which is ensemble-averaged over all particles in the field of
view. Figure~\ref{fig:msd-single} shows the MSD for one particular
soap film (solid symbols, refer caption of figure for details)
plotted against the lag time $\tau$. Also shown in the figure
is the corresponding MSD for diffusion in a 3D solution that
comprises the fluid layer of the soap film (open symbols). From the
figure, it is clear that the MSD is linear with respect to $\tau$,
indicating free diffusion. By comparing the two MSDs, it is also
evident that diffusion in the soap film (solid circles) is faster
than in the corresponding bulk solution (open circles). This makes
sense, as the Trapeznikov approximation states that the particle is
at an effective interface in contact with bulk air phases. Since
the air phase has a significantly lower viscosity than the fluid
layer, this speeds the diffusive motion of the particle. Finally,
Eqn.~\ref{Saffman} can be solved to estimate the effective surface
viscosity $\eta_{S}$ from the slope of the MSD.

\begin{figure} [tbhp]
\includegraphics[scale=0.38]{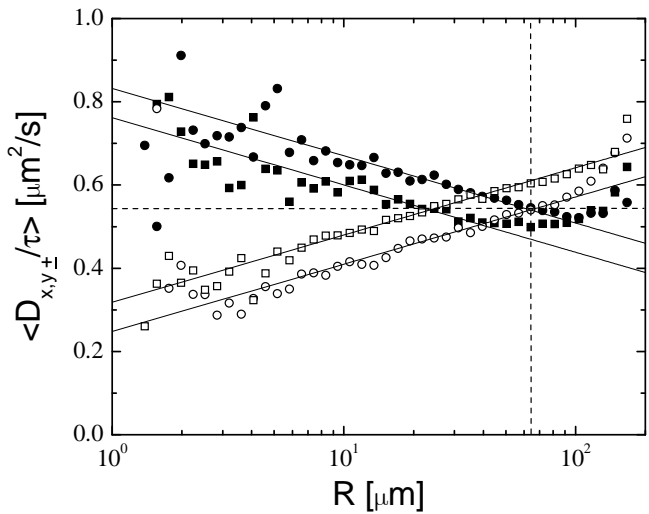}
\caption{\label{fig:Drr-tt-single} Correlation functions for the
same soap film as in Fig.~\ref{fig:msd-single}. Symbols are:
$\langle D_{x+}/\tau\rangle$, solid circles; $\langle
D_{x-}/\tau\rangle$, open circles; $\langle D_{y+}/\tau\rangle$,
solid squares; $\langle D_{y-}/\tau\rangle$, open squares. Solid
lines are fits to the data from Eqn~\ref{Diamant}, with
$B=0.54\mu$m$^2$/s (horizontal dashed line), $C=0.13$ and $L=64
\mu$m (vertical dashed line).}
\end{figure}

\begin{figure} [tbhp]
\includegraphics[scale=0.38]{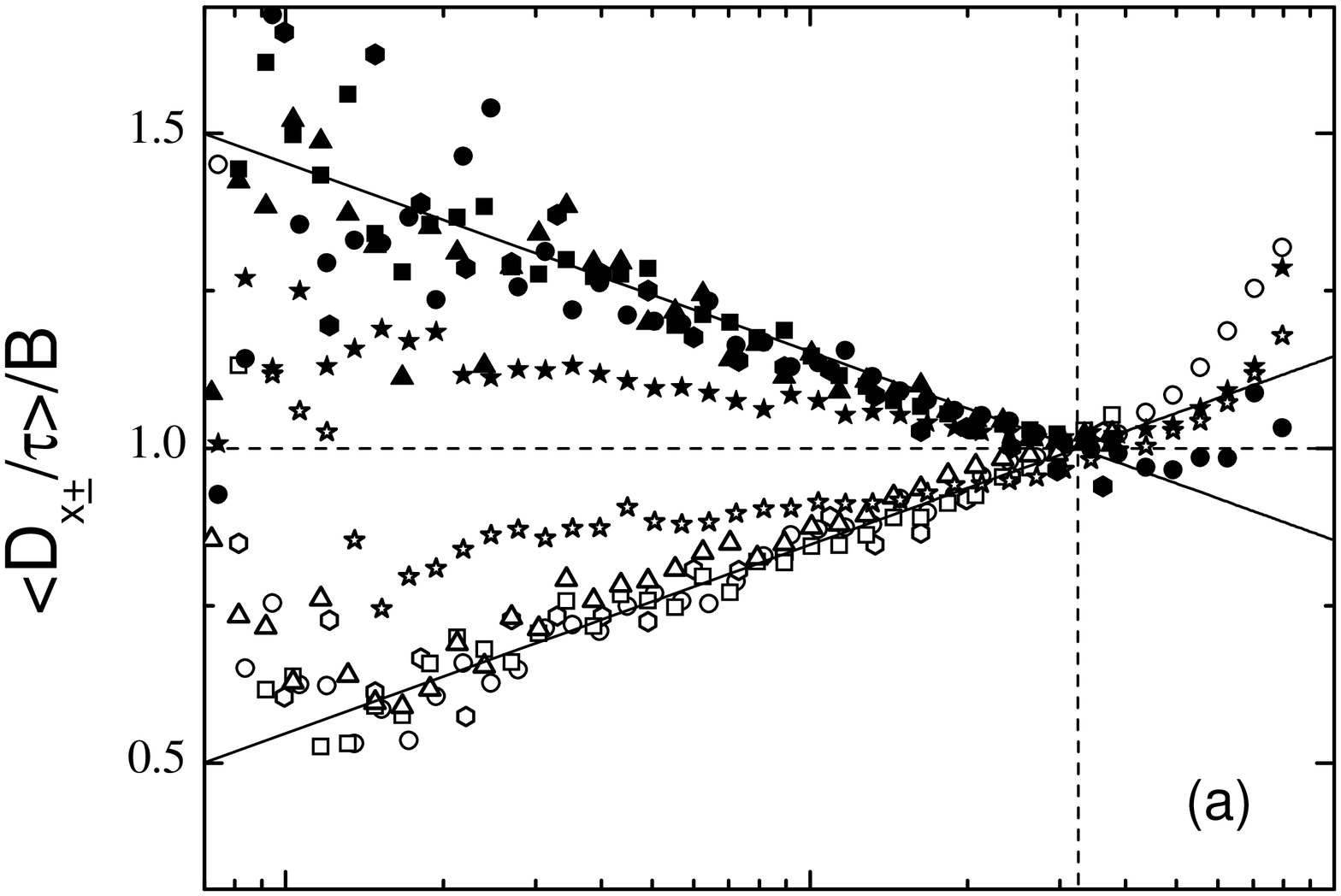}
\includegraphics[scale=0.38]{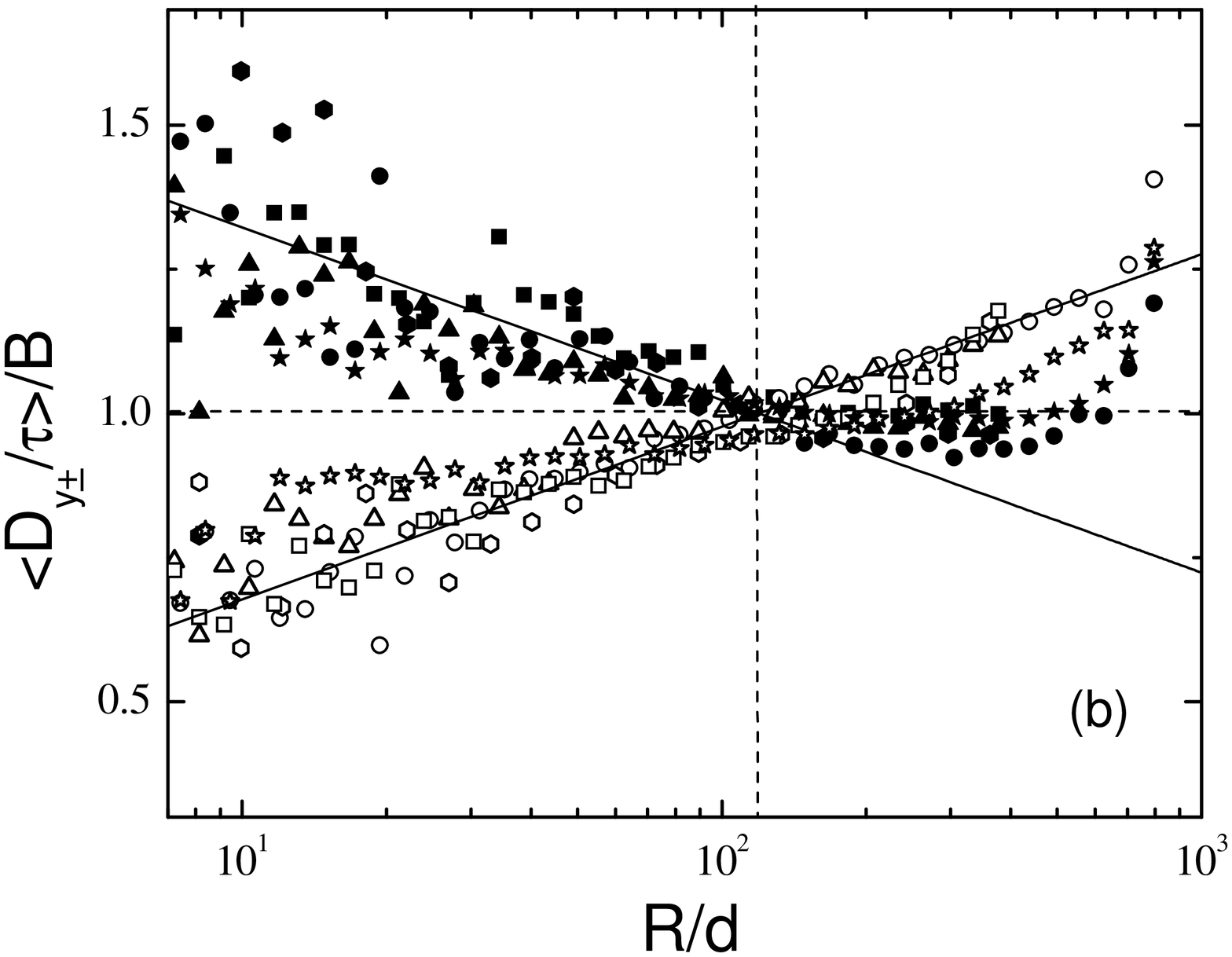}
\caption{\label{fig:Drr-scale} Scaled (a) longitudinal ($\langle
D_{x\pm}/\tau\rangle/B$) and (b) transverse ($\langle
D_{y\pm}/\tau\rangle/B$) correlation functions for five soap films
plotted against the scaled separation $R/d$. Solid symbols represent
correlated motion while open symbols represent anti-correlated
motion. Refer Table \ref{table1} for details about the soap films.
Symbols are: triangles, sample a; hexagons, sample c; squares,
sample d; circles, sample f; stars, sample i. Solid lines are fits
of the form $1\pm0.13~\text{ln}(325d/R)$ and $1\pm
0.13(\text{ln}(325d/R)-1)$ respectively.}
\end{figure}

After the effective surface viscosity has been determined, we
attempt to determine the flow fields in this soap film. This is done
by looking at the correlated and anti-correlated motions of pairs of
particles, as described in Sec.~\ref{tracking} of this
manuscript. The four correlation functions $\langle
D_{x+}/\tau\rangle$, $\langle D_{x-}/\tau\rangle$, $\langle
D_{y+}/\tau\rangle$ and $\langle D_{y-}/\tau\rangle$ are shown in
Fig.~\ref{fig:Drr-tt-single} as a function of particle separation
$R$. From the figure, it is clear that the correlation functions
split around a mean diffusion constant $B$, with rigid motions (+,
solid symbols) being favored and relative motions (-, open symbols)
opposed. Further, the correlation functions vary logarithmically as
a function of $R$, evidenced by the linearity of the data on a
log-lin plot. In fact, the data are well characterized by
Eqn.~\ref{Diamant}, where $B$, $C$ and $L$ are fitting parameters. A
visual way of determining the fit parameters is by noting that
$\langle D_{x+}/\tau\rangle=\langle D_{x-}/\tau\rangle=B$ when
$R=L$. Therefore, the dashed horizontal and vertical lines in the
figure indicate that $B=0.54 \mu$m$^2$/s and $L=64 \mu$m, while the
slope of the four correlations functions simply gives $C=0.13$.
These values have been used to fit the four correlation functions
with Eqn.~\ref{Diamant} in Fig.~\ref{fig:Drr-tt-single}. We should
note that the length scale determined above is smaller than the
field of view of the microscope, and is therefore not subject to
finite size effects; that is, it truly represents a cut-off length
scale for stresses in this particular soap film. Further, as $B$
has units of a diffusion constant, it can be related to the
self-diffusion of a single particle in the soap film (by replacing
$R=d$ in Eqn.~\ref{Diamant}). Finally, the constant $C = 0.13$
represents how quickly the flow fields decay in this particular soap
film.

\begin{table}
\caption{\label{table1} Material parameters for the nine soap films
described in this paper. $\eta_\text{{bulk}}$ (determined from
diffusivity measurements in bulk solutions) has an error of $\pm
5\%$, and values of $h$ and $d$ are certain to within $\pm 2\%$. The
uncertainties in $\eta_{\text{int}}$, derived from Eqn.~1 and 2, are
given in the brackets.}
\begin{ruledtabular}
\begin{tabular}{lccc}
$\eta_{\text{bulk}}$ [mPa$\cdot$s]&$h$ [nm]&$d$ [nm]&$h/d$\\
\hline
 a. 2.3&305&500&0.6 \\
 b. 3.0&640&500&1.3 \\
 c. 6.0&510&500&1.0 \\
 d. 10.0&1340&500&2.7 \\
 e. 25.0&1100&500&2.2\\
 f. 10.0&780&210&3.7\\
 g. 25.0&2184&210&10.4\\
 h. 30.0&2100&210&10.0\\
 i. 30.0&3000&210&14.3 \\
\end{tabular}
\end{ruledtabular}
\end{table}

\begin{figure} [tbhp]
\includegraphics[scale=0.35]{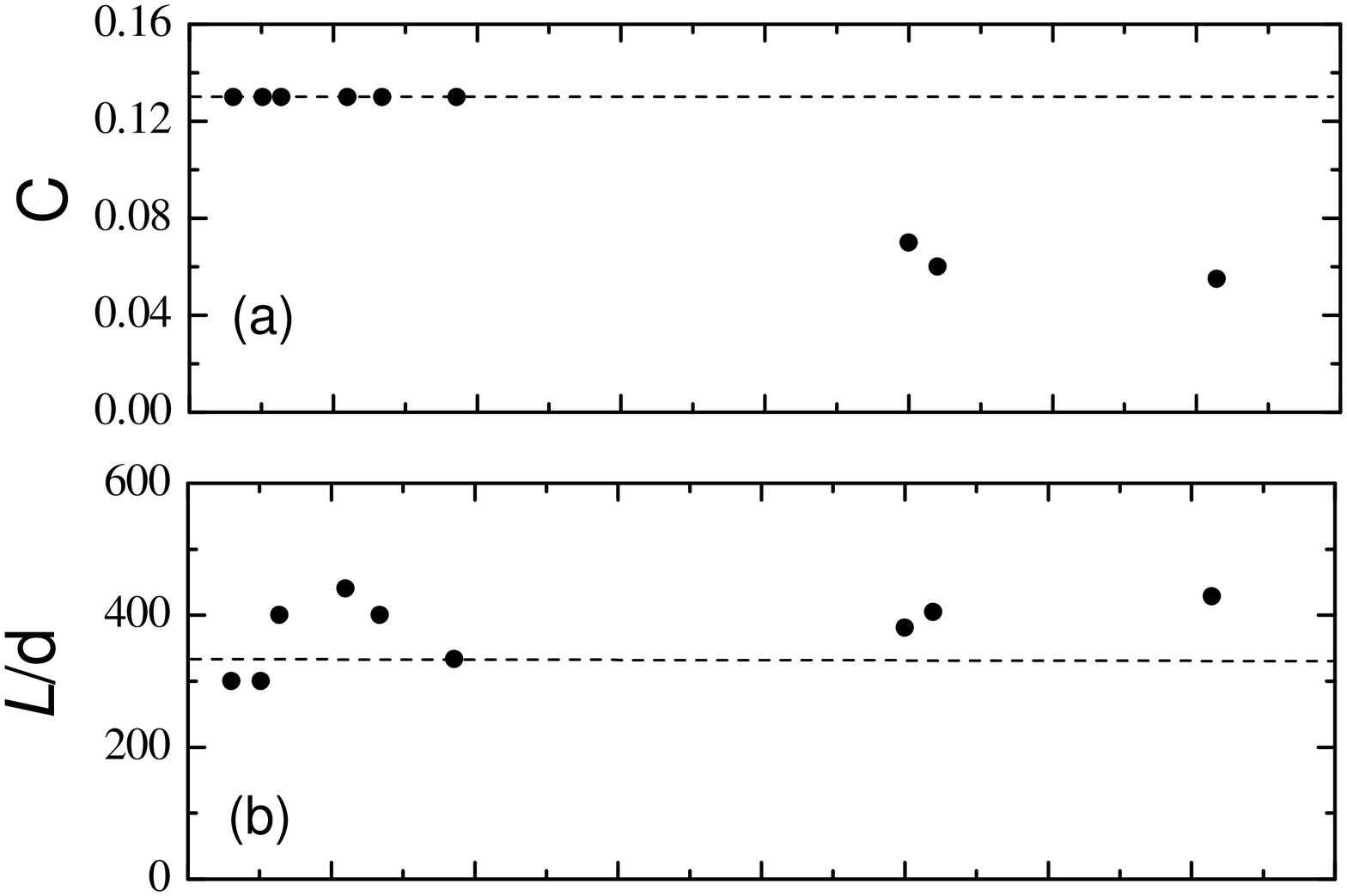}
\includegraphics[scale=0.35]{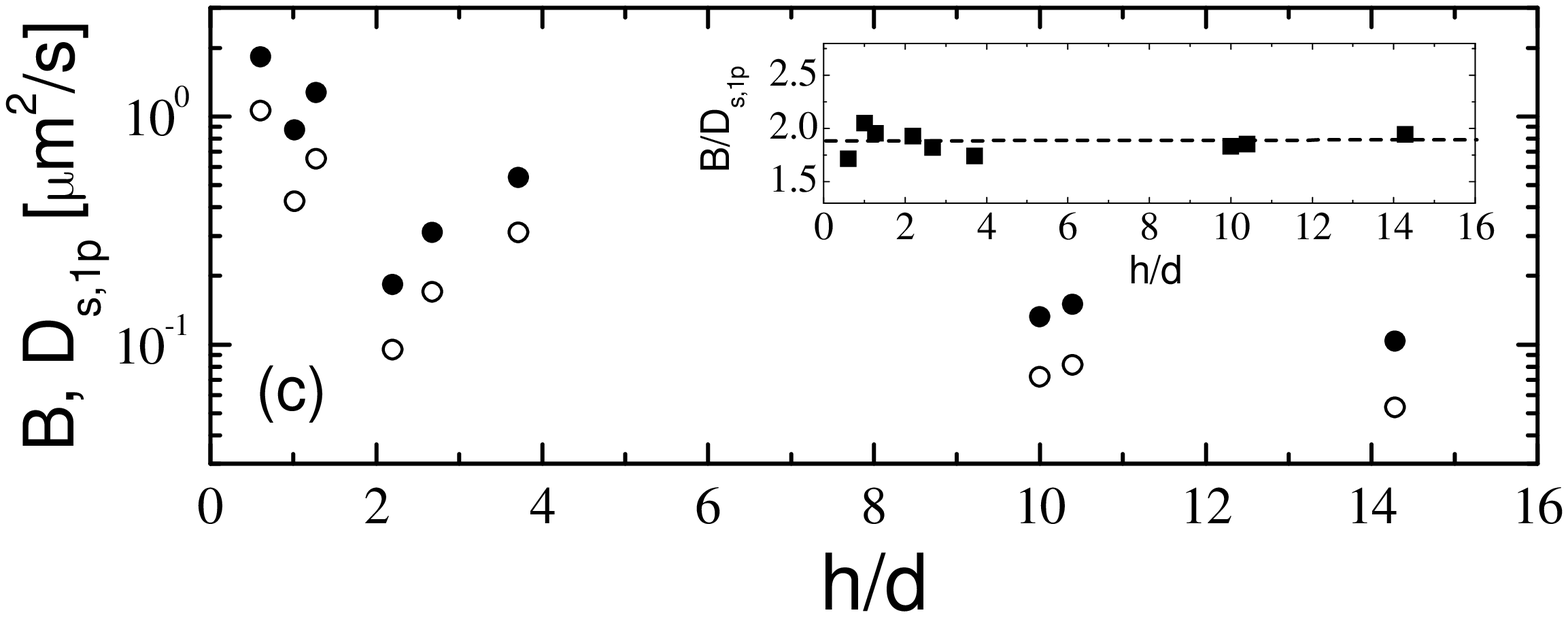}
\caption{\label{fig:fitparams} Fit parameters for nine soap films,
including the five shown in Fig.~\ref{fig:Drr-scale}, as a function
of $h/d$. (a) $C$; the dashed line indicates a constant value of
0.13 for all soap films with $h/d < 7\pm 3$.  (b) $L/d$; the
dashed line represents a value of 325, from the scaling of the
correlation functions shown in Fig.~\ref{fig:Drr-scale}. (c) $B$
(solid circles) compared to the one-particle diffusion constant
$D_{s,1p}$ (open circles). The inset shows the ratio $B/D_{s,1p}$ to
be nearly constant with a value of 1.9 $\pm$ 0.1.}
\end{figure}
To test the validity of the functional form of the correlation
functions, we measure and plot their values for different soap films
with a range of bulk viscosities, thickness and tracer particle
sizes. This is shown in Figs.~\ref{fig:Drr-scale}(a) and (b), where
we have non-dimensionalized the correlation functions by the scale
factor $B$ and the separation $R$ by the particle diameter $d$ for
five different soap films, including the one shown in
Fig.~\ref{fig:Drr-tt-single}. We find that for all soap films with
$h/d < 7$, the longitudinal correlation functions scale onto a
single curve, that is described by the equation $1\pm 0.13~\text{ln}
(325d/R)$. For the film where $h/d=14.3$ (stars), there is
significant deviation from the scaling, particularly relating to the
slope $C$ ($C=0.06$ instead of 0.13). Similarly, the transverse
correlation functions for the thin films can be well described by
the form $1\pm 0.13~(\text{ln} (325d/R)-1)$, while the thicker films
again deviate from the scaling (the transverse correlation functions
are noisier than their longitudinal counterparts, hence the
deviation is not as clearly visible). The dashed horizontal lines in
Fig.~\ref{fig:Drr-scale}(a) and (b) depict the splitting of the
normalized diffusion constants around a value of 1, while the
vertical lines are placed at values of $R/d =325$ and
$325/e~(=120)$ for the longitudinal and transverse correlation
functions respectively.

\begin{figure} [tbhp]
\includegraphics[scale=0.37]{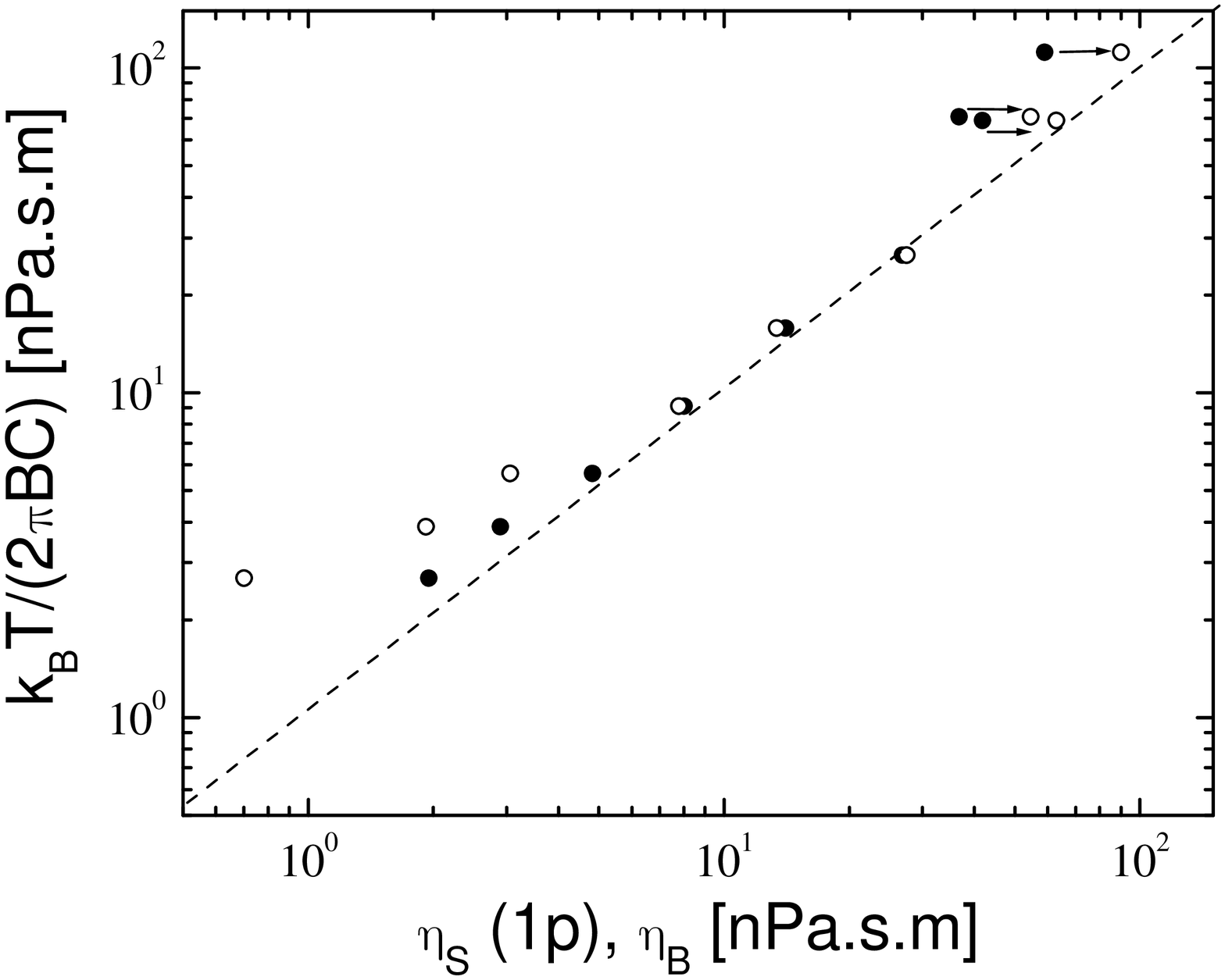}
\caption{\label{fig:b-vs-eta} The scaled fit parameter $k_{B}T/(2\pi
BC)$ compared to the one-particle effective surface viscosity
$\eta_{S}$ (solid circles) and $\eta_{B}$ (open circles). The dashed
line indicates equality between the two quantities.}
\end{figure}

The fitting parameters from the five soap films described above, as
well as four additional ones are shown in Fig.~\ref{fig:fitparams}.
The fit parameter $C$ is plotted against $h/d$ in
Fig.~\ref{fig:fitparams}(a), where we see that $C\approx 0.13$ for
all films with $h/d < 7\pm 3$, while $C\approx 0.06$ for thicker
films. On the other hand, the parameter $L/d$ shown in
Fig.~\ref{fig:fitparams}(b) is nearly constant for \emph{all} soap
films. Finally, in Fig.~\ref{fig:fitparams}(c) we plot both $B$ and
$D_{s,1p}$ as a function of $h/d$. Neither the fit parameter nor the
one-particle diffusion constant show any discernible trend with
$h/d$; however, judging by their equidistant separation on a
logarithmic scale, they seem to be correlated together in some
fashion. Indeed, plotting $B/D_{s,1p}$ against $h/d$, we find that
$B/D_{s,1p} \approx 1.9 \pm 0.1$ for all soap films. This is
perhaps
surprising, as $D_{s,1p}$ is a local measurement determined from the
trajectory of single particles, while $B$ is obtained from the
correlated motion of pairs of particles.

These parameters can be understood by considering various
potentially relevant quantities which all have units of surface
viscosity.  The first quantity is $\eta_B = \eta_{\text{bulk}}h$.
This product shows up in the theory (Eqns.~\ref{diLeonardo-theory},
\ref{Leonardo-Haim} and \ref{Diamant}), and is an appealing
quantity because it depends on two physical parameters that
are easy to measure.  The second quantity is the effective
surface viscosity $\eta_S$ as determined from the one-particle
measurements (Eqn.~\ref{Saffman}).  The third quantity comes
from the Trapeznikov approximation (Eqn.~\ref{trapeqn}),
which states that the entire soap film can be considered
as an effective interface with a larger surface viscosity
$\eta_T = \eta_{\text{bulk}}h + 2\eta_{\text{int}} = \eta_B +
2\eta_{\text{int}}$, thus suggesting that $\eta_T$ is more
relevant than the first quantity $\eta_B$.  The fourth quantity
is the combination $k_{B}T/(2\pi BC)$, based on the two-particle
correlations.  From examining the theoretical expressions for the
correlation functions described in Eqns.~\ref{diLeonardo-theory},
\ref{Leonardo-Haim} and \ref{Diamant}, we see that $B=2k_{B}Tb$
and $C=1/(4\pi\eta_{\text{bulk}}hb)$, so that $k_{B}T/(2\pi BC)
= \eta_B$.  Thus, testing this equality is a test of that theory.
In particular, the contribution to $\eta_T$ from $\eta_{\text{int}}$
has not been included in the expression for $C$, and so it is
possible that the theory needs to be modified.  We note that the
fluid layer in the soap film studied by di Leonardo \emph{et al.}
\cite{diLeonardo1} consisted primarily of glycerol, and presumably
$\eta_{\text{int}}$ was irrelevant for their system, that is,
$\eta_B \approx \eta_T \gg \eta_{\text{int}}$. This is not true
for our soap films, as we have situations where $\eta_B \sim
\eta_{\text{int}}$ as well as $\eta_B \gg \eta_{\text{int}}$.

To test these conjectures, we compare these three quantities in
Fig.~\ref{fig:b-vs-eta}, where we plot the scaled quantity
$k_{B}T/(2\pi BC)$ against the measured one-particle effective
surface viscosity $\eta_S$ (solid circles) and
$\eta_B$ (open circles).  For thin films with low
surface viscosities (lower left corner of the graph), we see the
solid symbols for $\eta_S$ are to the right of the open
symbols for $\eta_B$, in agreement with the Trapeznikov
approximation.  From the difference in these two quantities, one
can extract the surface viscosity $\eta_{\text{int}}$ which we
have done previously for these data, finding $\eta_{\text{int}}
= 0.97 \pm 0.55$ nPa$\cdot$s$\cdot$m \cite{prasad3}.  However,
for thicker films with higher surface viscosities (top right
corner of Fig.~\ref{fig:b-vs-eta}), we see the opposite is
true; the effective surface viscosities $\eta_S$ (solid circles) are
lower than $\eta_B$ (open circles),
showing that
Saffman's equation (Eqn.~\ref{Saffman}) measures a viscosity
that is too low.  This occurs even though the thick films still
behave as a 2D fluid according to their correlation functions,
in other words, Eqns.~\ref{diLeonardo-theory}, \ref{Leonardo-Haim}
and \ref{Diamant} still describe the two-particle correlations,
albeit with different parameters.

The comparison with $k_{B}T/(2\pi BC)$ in Fig.~\ref{fig:b-vs-eta}
bolsters these conclusions.  For thin films, $\eta_S$
matches better with $k_{B}T/(2\pi BC)$ (comparison of solid symbols
with the dashed line, lower left of Fig.~\ref{fig:b-vs-eta}).
For thicker films, $\eta_B$ matches better
(comparison of open symbols with the dashed line, upper right of
Fig.~\ref{fig:b-vs-eta}).  Our data suggest that in the theory
leading to Eqns.~\ref{diLeonardo-theory}, \ref{Leonardo-Haim},
and \ref{Diamant}, we should replace $\eta_B = \eta_{\text{bulk}}h$
with $\eta_T$; this minor correction
would improve the results for thin, less viscous films, and be
negligible for thicker viscous films such as those studied in
Ref.~\cite{diLeonardo1}.  The disagreement between $k_{B}T/(2\pi
BC)$ and $\eta_S$ shows the breakdown of the
applicability of Saffman's equation to extract a useful surface
viscosity from one-particle data, for situations with $h/d > 7$,
and in fact coincides with a breakdown of the equality $\eta_S =
\eta_T$ \cite{prasad3}.

To summarize, for all films we have $k_{B}T/(2\pi BC) = \eta_T$ and
for thin films we additionally have $\eta_S = \eta_T$. The
approximation $\eta_B \approx \eta_T$ is mathematically obvious for
thick films, given the definitions of these two surface viscosities.
It is worth noting that our prior work suggests that it is necessary
to measure $\eta_S$ (in thin films) to be able to determine
$\eta_T$, as the viscosity of the surfactant layers
$\eta_{\text{int}}$ is usually not known ahead of time
\cite{prasad3}; this current work suggests that measuring
$k_{B}T/(2\pi BC)$ is an additional method to obtain $\eta_T$ and
thus $\eta_{\text{int}}$.

Using these insights, we return to the data shown in
Fig.~\ref{fig:fitparams}.  For low values of $h/d$, $C$ is constant,
as shown in panel (a); meanwhile, $B$ drops, as shown in panel (c).
This suggests that the mobility $b$ scales as $b \sim \eta_T^{-1}$
for thin films.  Indeed, this explains the correlations of $B$ and
$D_{s,1p}$ seen in Fig.~\ref{fig:fitparams}(c) for thin films, as
$D_{s,1p} \sim 1/\eta_T$ would be expected \cite{prasad3}.  However,
at thicker films, $C$ decreases, suggesting that $b$ grows faster
than $\eta_T^{-1}$; the mobility of particles increases. This is
consistent with Saffman's equation incorrectly yielding a surface
viscosity $\eta_S$ that is too low for the thicker films.

We also note that the parameter $L/d$ is nearly constant for
\emph{all} soap films. This is very surprising, as the logarithmic
cut-off in the decay of the correlation functions can arise from
three possible factors: 1) finite size of the film; (2) inertial
effects; and (3) viscous drag on the soap film from the bulk air
phases. The size of the frame used to house the film is of order
$\sim$ 1 cm, which is too large to explain the values of $L$
obtained from fits to the correlation functions. Inertial effects
arise at length scales given by $L=\eta_{\text{bulk}}/\rho U$, where
$\rho$ is the fluid density and $U$ is a typical probe particle
speed. This gives length scales of order $\sim$ 1m, which is again
too large to explain our fit-derived length scales. Finally, the
viscous cut-off length scale is given by
$L=\eta_{s,\text{eff}}/\eta_{\text{air}}$. Clearly, as can be seen
from Fig.~\ref{fig:b-vs-eta}, this length scale must increase as the
effective surface viscosity increases. Therefore, the near-constant
value of $L/d$ for soap films over a range of $\eta_{s,\text{eff}}$
remains a mystery.  Note that none of these potential origins for
$L$ would predict any dependence on the particle size $d$; that
is, they might predict a film-independent $L$ but not a constant $L/d$ as
we find (from data including two different particle sizes $d$
differing by a factor of 2.4).

One additional particle-based explanation for the constant value
of $L/d$ could be capillary effects. Capillary effects arise from
deformation of the interface by the particle, with an energy gain
obtained when two particles stick to each other.  However, it isn't
obvious that this moderately short range interaction would lead
to such a very long length scale as we observe ($L/d \sim 325$).
We leave the reasons for the constant value of $L/d$ as a matter
for future research.

\section{Conclusion}

We have used the technique of two-particle microrheology to
characterize the flow fields in soap films of varying thickness
$h$, where $0.6\leq h/d \leq 14.3$ (based on probe diameter
$d$). In particular, we determine the `eigenmobilities'
of thermally correlated motions of probe particles in these
soap films.  These eigenmobilities consist of correlated motion
parallel and perpendicular to the lines joining the centers of
pairs of particles, and rigid and relative motion as well. The
eigenmobilities are found to split around a mean value and decay
logarithmically as a function of particle separation for all soap
films. The flow fields of all films we observe are well described by
theoretical models.  A study of the fit parameters shows that thin
films have a simple behavior, where a surface viscosity $\eta_T$
predicted by Trapeznikov over 50 years ago correctly describes
both one-particle and two-particle motions \cite{Trapeznikov1}.
In fact, for the thinnest films, our results suggest that prior
theoretical work should be modified slightly to use $\eta_T$
\cite{diLeonardo1}.  In our work we also a transition
from `pure-2D' to `3D-influenced' behavior at $h/d = 7 \pm 3$.
For thicker films, two-particle correlations still follow the
predicted form for a quasi-2D fluid, with surface viscosity
$\eta_T$.  However, one-particle motion is faster than expected,
and other significant deviations from the thin-film behavior are
noted.  The results of our study on soap films can have important
consequences for other 2D systems, including but not limited to
microfluidic flow in confined geometries, protein and surfactant
monolayers at an air water interface and the cell membrane.

Funding for this work was provided by the National Science
Foundation (DMR-0804174) and the Petroleum Research Fund,
administered by the American Chemical Society (47970-AC9). We
thank J. Gallivan for use of the spectrophotometer, and H.~Diamant,
H.~A.~Stone, and E.~van Nierop for helpful discussions.


\end{document}